\documentclass[12pt]{article}

\pdfoutput=1

\usepackage{amsmath,amsfonts}
\usepackage{amssymb}
\usepackage{booktabs}
\usepackage[usenames,dvips]{color}
\usepackage{graphicx}
\usepackage{comment}
\usepackage{fullpage}

\graphicspath{ {figs/} }

\makeatletter
\@addtoreset{equation}{section}
\makeatother

\definecolor{lightgray}{rgb}{0.7,0.7,0.7}

\newcommand{\be}{\begin{equation}}
\newcommand{\ee}{\end{equation}}
\newcommand{\bea}{\begin{eqnarray}}
\newcommand{\eea}{\end{eqnarray}}

\newcommand{\tr}{\operatorname{tr}}

\newcommand{\nn}{\nonumber}

\renewcommand{\sl}{/\!\!\!}

\newcommand{\ms}{{\mathstrut}}

\begin{document}

\title{Natural Fermion Hierarchies from  Random Yukawa Couplings}
\author{Gero von Gersdorff\\
\normalsize Pontif\'icia Universidade Cat\'olica, Rio de Janeiro, Brazil}
\date{}

\maketitle
\abstract{The Standard Model of particle physics requires Yukawa matrices with eigenvalues that differ by orders of magnitude.
We propose a novel way to explain this fact without any small or large parameters. The mechanism is based on the observation that products of matrices of random order-one numbers have hierarchical spectra. The same mechanism can easily account for the hierarchical structure of the quark mixing matrix.}

\newpage

\section{Introduction}

One of the great unsolved puzzles of the Standard Model (SM) concerns the large mass hierarchies of the quarks and leptons.
For the up quark, down quark and charged leptons sectors, one finds respectively $m_t/m_u\sim 6.9\times 10^4$, $m_b/m_d\sim 8.7 \times 10^2$ and $m_\tau/m_e\sim  3.4\times 10^3$.
It appears quite unnatural that such large hierarchies arise purely by chance from generic Yukawa matrices.

The most popular solutions to this problem employ an order parameter $\epsilon\sim 0.1$ in terms of which the  Yukawa couplings scale as (up to order one coefficients)
\be
(Y_u)_{ij}\sim \epsilon^{q_i+u_j}\,,\qquad (Y_d)_{ij}\sim \epsilon^{q_i+d_j}\,,\qquad (Y_e)_{ij}\sim \epsilon^{\ell_i+e_j}
\label{eq:FN}
\ee
where $q_i$, $u_i$, $d_i$, $\ell_i$, and $e_i$ are real positive numbers of order unity.  Such relations can be given a convincing UV description in terms of spontaneously broken $U(1)$ symmetries \cite{Froggatt:1978nt} or in terms of wave-function localization in extra dimensions \cite{Grossman:1999ra,Gherghetta:2000qt,Huber:2000ie}. If flavor violating new physics (NP) is present at the TeV scale, these models typically feature somehow suppressed flavor changing neutral currents (FCNC) (as opposed to a generic flavor structure), but nevertheless are tightly constrained, in particular by data on $K\bar K$ mixing and $\mu\to e\gamma$ decays (see, for instance, Ref.~\cite{vonGersdorff:2013rwa} for a review of flavor bounds in extra dimensions.).

In this short note we suggest a new  mechanism that provides a natural explanation of the fermion mass hierarchies of the SM. Instead of invoking an order parameter, the mechanism is based on completely random couplings of order one. Due to the properties of the probability distributions for the effective Yukawa couplings  extreme ratios of eigenvalues become completely common, and mass hierarchies are hence the rule rather than the exception. 
We will refer to this mechanism as the {\em stochastic hierarchy mechanism}. The peculiar almost-diagonal nature of the Cabbibo-Kobayashi-Maskawa (CKM) matrix can also easily explained with our approach, as the alignment/misalignment of the up and down Yukawa matrices can be controlled by some judicious choice of the matrix products.

The "randomness" hypothesis has previously been considered in the context of neutrino mixing \cite{Hall:1999sn} and is commonly referred to as neutrino anarchy. Moreover, anarchic perturbations to a hierarchical model were investigated in Ref.~\cite{Rosenfeld:2001sc}, and  in the context of supersymmetry it was found that some hierarchies can also arise stochastically \cite{Altmannshofer:2014qha}. 
Furthermore, in models giving rise to Yukawa couplings of the type Eq.~(\ref{eq:FN}), the unknown $\mathcal O(1)$ numbers multiplying the suppression factors are often considered to be stochastic as well. 
However, the goal of the present manuscript is the generation of the large (charged) fermion hierarchies "out of nothing", i.e., from purely $\mathcal O(1)$ random couplings, which to the best of our knowledge has never been achieved in the literature.

\section{Products of random matrices}

As a warm-up to the study of more realistic models, let us imagine that the Yukawa matrices  of the SM are given by a product of individual "proto-Yukawa" matrices:
\be
Y= Y_1 Y_2\cdots Y_N\,.
\label{eq:randprod}
\ee
This is a simplified version of the effective Yukawa coupling arising from the explict models constructed in the next section. However, it will be sufficient for the illustration of the most important features of the stochastic hierarchy mechanism.

In the absence of an explicit UV model for  the proto-Yukawas $Y^i_{ab}$, our best guess is that they are random $\mathcal O(1)$ numbers. We will assume that they are real and follow some  "base distribution" (or "prior") with mean zero and variance $\sigma^2$. A simple and natural choice is the uniform distribution with
\be
-1<Y^i_{ab}<1\,,
\ee
which has variance $\sigma^2=1/3$, but many of our results below are valid for any other symmetric prior. 

We would like to find the probability distribution for the largest mass hierarchies in the matrix  $Y_{ab}$,
\be
h\equiv \max |y_i|/\min|y_i|\,,
\label{eq:hdef}
\ee
where $y_i^2$ are the three eigenvalues of $YY^T$.
In theory, the analytical calculation of these distributions is straightforward: one substitutes one of the matrix elements by $h$ and marginalizes (integrates) over the remaining ones.
Unfortunately, this calculation is obstructed by the resulting complicated region of integration, and one has to result to numerical simulations.
However, some aspects of these distributions can be computed analytically and allow for a rough understanding of the mechanism.

Rather than calculating the full distributions we will focus on their lowest moments. 
 For instance, for mean and (co)variance of  $Y_{ab}$ one easily finds 
\be
\langle Y_{ab} \rangle=0\,,\qquad \langle Y_{ab} Y_{cd} \rangle=\frac{1}{N_f} \left(N_f\,\sigma^{2}\right)^N \delta_{ab,cd}\ ,
\ee
where $N_f$ denotes the number of families. Note that the correlations between different matrix elements vanish. For higher moments this is no longer true and the matrix elements are in fact statistically dependent.
Interestingly, for $N_f=3$ and flat priors, the variance is independent of $N$ and  equals the one of the flat prior ($\sigma^2=1/3$). 

Let us now examine the distribution for the \emph{determinant} of $Y$. One finds
\be
\langle \det Y\rangle=0\,,\qquad \langle (\det Y)^2 \rangle =\left(N_f!\,\sigma^{2N_f}\right)^N.
\label{eq:det}
\ee
The power of $N$ in the expression for the variance is a simple consequence of the determinant multiplication theorem which causes the integrals over $(Y_i)_{ab}$ and $(Y_j)_{ab}$ to factorize.
For $N_f=3$ and the flat prior, the variance is given by $(2/9)^N$ and hence 
the determinant tends to be very suppressed despite the fact that the matrix elements are typically of $\mathcal O(1)$. 
At first this seems at odds with the fact that the covariance matrix is diagonal.
However, correlations of higher moments do not vanish, and it is those that make these cancellations possible.

To get a feeling for the typical size of eigenvalues, we can 
compute the mean of $\tr YY^T$.
One gets
\be
\langle \tr Y Y^T \rangle = N_f\left(N_f\, \sigma^2\right)^N.
\label{eq:tr}
\ee
Let us focus first on the flat uniform prior. For $N_f=3$, Eq.~(\ref{eq:tr}) becomes independent of $N$, and  from Eq.~(\ref{eq:tr}) and (\ref{eq:det}) we get the estimate
\be
y_1^2+y_2^2+y_3^2\sim 3\,, \qquad |y_1 y_2 y_3| \lesssim \left(\frac{ 2}{9}\right)^{N/2}.
\label{eq:est1}
\ee
Thus, at least one eigenvalue has to be of $\mathcal O(1)$, and the geometric mean of the other two is suppressed.
Ordering $y_1,y_2\ll y_3$, one gets from Eq.~(\ref{eq:est1}) the estimate
\be
\frac{
(y_1y_2)^\frac{1}{2}}{y_3}\lesssim \frac{1}{3^\frac{3}{4}}\left(\frac{ 2}{9}\right)^{N/4}\sim 0.43\times 0.68^N\,.
\label{eq:est2}
\ee
Interestingly, this ratio is actually independent of $\sigma^2$ and hence valid for any prior, even though the size of the largest eigenvalue is no longer independent of $N$ and can be both supressed or enhanced, depending on wether $\sigma^2$ is smaller or larger than $1/3$.\footnote{
The behaviour of hierarchies of random matrices at large $N_f$ (rather than $N$) has been studied previously in the  context of neutrino masses \cite{Bai:2012zn}. } 

\begin{figure}
\centering
\includegraphics[width=8cm]{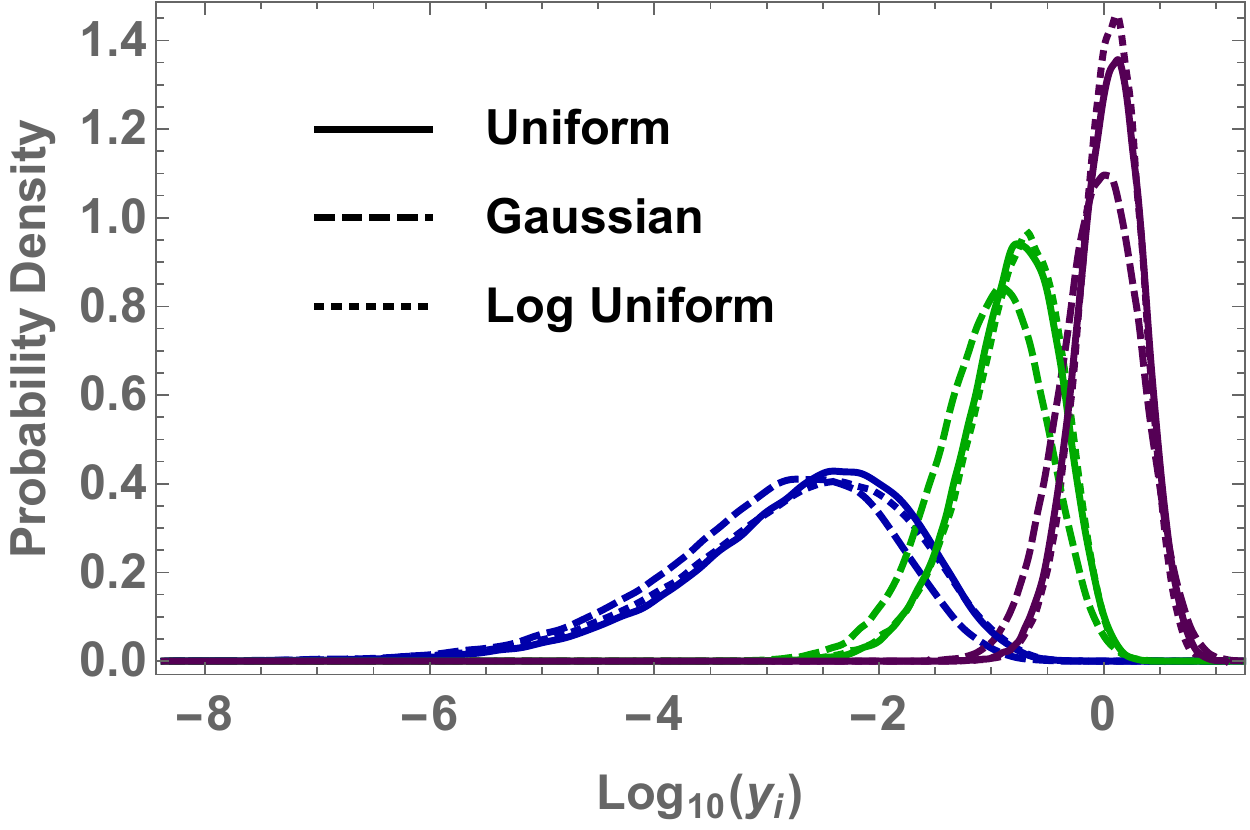}
\caption{Distribution of the small, medium, and large eigenvalues for $N=7$ for different priors. }
\label{fig:toymodel}
\end{figure}

Finally, we comment that there is typically also a hierarchy between the lighter two eigenvalues. To show this, we would have to examine the distributions of other quantities (such as the principal minors), but we will not go into this much detail here. Instead we show in Fig.~\ref{fig:toymodel} the simulated distributions of the $N_f=3$ eigenvalues for the case $N=7$, from which the hierarchical spectrum $y_1\ll y_2\ll y_3$ is quite evident. We also provide, for comparison, the cases of a Gaussian prior and a log-uniform prior, showing excellent prior-independence.\footnote{All priors are chosen to have the same variance. As already mentioned, the absolute scale of the eigenvalues depends on the variance, but the hierarchies do not.}
We also find empirically 
\be
\frac{y_3}{y_2}<\frac{y_2}{y_1}
\label{eq:partialhier}
\ee
in about $80\%$ of the cases, roughly independent of $N$, which shows that up-like and lepton-like spectra are more common than  down-like ones.

Next we would like to examine the CKM mixing matrix. It also has a hierarchical structure, with diagonal elements $\sim 1$ and  more  suppressed entries the more one moves away from the diagonal. The CKM matrix is the ratio of the left handed rotation matrices 
\be
V_{\rm CKM}=V^{\phantom\dagger}_{u_L}\!V^\dagger_{d_L}\,,
\ee 
where $V_{u_L}$ and $V_{d_L}$ are 
the unitary matrices diagonalizing the combinations $Y^uY^{u\,\dagger}$ and $Y^dY^{d\,\dagger}$ respectively.
Therefore, in order to obtain an approximately diagonal CKM matrix, the two left handed rotations have to be similar, or equivalently, the up and down Yukawa couplings have to be roughly aligned. 
This can easily be achieved as follows. In addition to the physical Yukawas being products of several matrices, we stipulate that some of these matrices are the same (we will see in the next section how this can be achieved in a model):
\be
Y^d=Y^q_{1}\cdots Y^q_{N_q}\, Y^{d\ms}_1 \cdots  Y^{\ms d}_{N_d}\,,\qquad 
Y^u=Y^q_{1}\cdots Y^q_{N_q}\, Y^{u\ms}_1 \cdots  Y^{u\ms}_{N_u}\,.
\label{eq:CKMmodel}
\ee
Note that the first $N_q$ factors are common, and it is this common factor that guarantees a certain degree of (left-handed) alignment. 
Without this factor, there would be no alignment and the CKM matrix becomes democratic. 
To see why the CKM matrix is hierarchical, it is instructive to  diagonalize the respective factors 
\be
Y^d=U_q \hat Y^q \tilde U_q^\dagger  \tilde U^\ms_d \hat Y^{\ms d}U_d^\dagger\,,\qquad 
Y^u=U_q \hat Y^q \tilde U_q^\dagger  \tilde U^\ms_u \hat Y^{\ms d}U_u^\dagger\,,
\label{eq:CKMmodel}
\ee
Here, $\hat Y^{q,u,d}$ are diagonal and hierarchical (according to the numbers $N_q$, $N_u$, and $N_d$). It is important to stress that the unitary matrices  $\tilde U_{q,u,d}$ and $U_{q,u,d}$ are not hierarchical. Making a gauge-invariant change of basis with the matrices $U_{q,u,d}$, we obtain the structure
\be
(Y'^d)_{ij}\sim \hat y^q_i\hat y^d_j\,,\qquad  (Y'^u)_{ij}\sim \hat y^q_i\hat y^u_j\,,\qquad 
\ee
where we have not written $\mathcal O(1)$ numbers arising from the $\tilde U_{q,u,d}$. In this new basis, the matrices take the familiar Frogatt-Nielsen (FN) form, Eq.~(\ref{eq:FN}). As is well known, for hierarchical $\hat y^{q,u,d}$, the CKM matrix scales as \cite{Froggatt:1978nt}
\be
(V_{\rm CKM})_{ij}\sim \exp\left( -\left|\log\, {\hat y^q_i}/{\hat y^{q}_j}\right|\right)\,.
\ee 
(again up to $\mathcal O(1)$ numbers), showing that the CKM hierarchy is determined entirely in terms of the hierarchies of the eigenvalues $\hat y^q_i$ of the common factor.

\begin{figure}
\centering
\includegraphics[width=8cm]{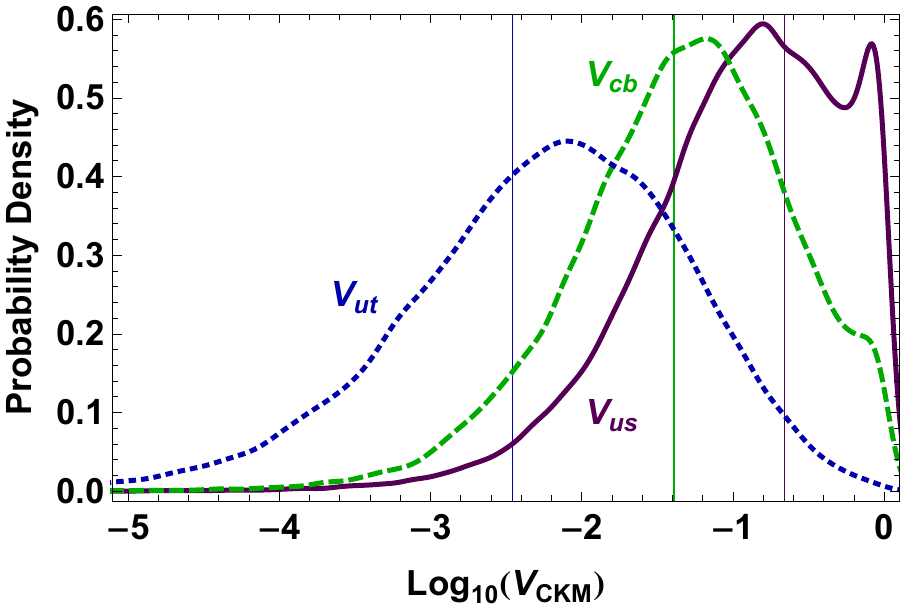}
\caption{Distribution of CKM matrix elements for $N_q=5$, $N_d=0$, $N_u=5$. The experimental values are indicated as vertical lines.}
\label{fig:CKM}
\end{figure}

We can choose the numbers $N_u$, $N_d$ and $N_q$ in order to maximize the probabilities to achieve SM-like values for the masses and mixings.\footnote{An additional global suppression factor can be introduced for the down sector, mimicking the effect of large $\tan\beta$ in supersymmetry. }
We have simulated the distributions for $V_{\rm CKM}$, drawing from flat priors for all the matrices involved in Eq.~(\ref{eq:CKMmodel}), for the case $N_q=5$, $N_d=0$, and $N_u=5$. In this simulation we have made the additional selection $y_3/y_2$ greater (smaller) than $y_2/y_1$ for the down (up) sector respectively, as occuring in the SM. The efficiency of this cut is about 0.13, which is a bit lower than the    value $0.8\times 0.2=0.16$ to be expected from the considerations around Eq.~(\ref{eq:partialhier}), because the eigenvalues in the up and down sector become mildly correlated. One can see clearly that the true CKM angles appear in the bulk of the distributions and are hence at their natural values.

\section{An explicit model}

Motivated by the observations of the previous section we now move on to construct a model with effective Yukawa couplings given by products of matrices. 
Consider replacing the SM (say up-type) Yukawa interaction by the Lagrangian
\bea
\mathcal L_u&=&\sum_{i=1}^{N_q}\bar Q_i(\sl p+M^q_i)Q_i-(\bar Q_iK^q_iQ_{i+1}+h.c.)\nn\\
&&+\sum_{i=1}^{N_u}\bar U_i(\sl p+M^u_i)U_i-(\bar U_iK^u_iU_{i+1}+h.c.)\nn\\
&&-(\bar Q_1\tilde H\,Y^u_0 U_{1}+h.c.)\,,
\label{eq:model}
\eea
where $Q_{N+1}=q_L$ and $ U_{N+1}\equiv u_R$ are chiral fields, the remaining $U_i$ and $Q_i$ are vector-like quarks, and $H$ is the SM Higgs field.
The masses $M_i$ are hermitian and the mass mixings $K_i$ arbitrary 3$\times$3 matrices. 
Lagrangians of this kind are familiar from discretizations of extra dimensions \cite{ Hill:2000mu} and composite Higgs models with partial compositeness \cite{Panico:2015jxa}. They have recently been reconsidered in the so-called clockwork mechanism \cite{Giudice:2016yja}. In contrast to these constructions, here no small parameter will be needed.~\footnote{
Our Lagrangian also resembles somewhat the model presented in Ref.~\cite{Giudice:2008uua} (Sec.~5) which also uses vectorlike fermions. In that model a judicious choice of the proto-Yukawa couplings ensures that after integrating out the heavy fermions some of the SM fermions only couple to higher powers of the Higgs field, creating the observed hierarchies. Here all fields have linear couplings to the Higgs, and the hierarchies appear via products of matrices, as explained below.}
Let us briefly comment on how one can achieve that the fermions $U_i$ ($Q_i$) only couple to $U_{i+ 1}$ ($Q_{i+ 1}$). One possibility is to promote the 
spurions $K_i$ and $M_i$ to physical fields that obtain vacuum expectation values via some mechanism. in fact, for vanishing couplings a large chiral symmetry (with a $U(3)_L\times U(3)_R$ at each site) emerges. The nearest neighbor interaction can then be achieved by  introducing physical fields $M_i$ and $K_i$ only in the bifundamentals of  $U(3)_{L,i}\times U( 3)_{R,i}$ and $U(3)_{L,i}\times U(3)_{R,i+1}$ respectively.

 We can  integrate out the  $U_i$ nad $Q_i$, yielding the effective Lagrangian 
\be
\mathcal L'_u=
\bar q_L\,\delta Z^q_{N_q} \sl p\, q_L
+\bar u_R\,\delta Z^u_{N_u} \sl p\, u_R
-\tilde H\bar q_L (\tilde Y^q_{1}\cdots \tilde Y^q_{N_q})^\dagger Y^u_0(\tilde Y^u_1\cdots \tilde Y^u_{N_u}) u_R+h.c.\,,
\label{eq:Leff}
\ee
with the recursively defined matrices
\be
\delta Z^\ms_k=\tilde Y_k^\dagger (1+\delta Z^\ms_{k-1}) \tilde Y^\ms_k\,,\qquad \delta Z^\ms_1=\tilde Y_1^\dagger\tilde Y_1^\ms\,,
\ee
and
\be
\tilde Y^\ms_k=(M^\ms_k-K_{k-1}^\dagger \tilde Y^\ms_{k-1})^{-1}K^\ms_k\,,\qquad  \tilde Y^\ms_1=M_1^{-1}K^\ms_1\,.
\ee
We stress that the effective Lagrangian Eq.~(\ref{eq:Leff}) is exact as long as none of the (true) masses of the heavy fields are at or below the electroweak scale. The expressions for $N>1$ quickly get quite complicated. However, the recursive definitions are well suited for numerical simulations, and in particular are much easier to handle than the full diagonalization of the mass matrix.

The down-type Yukawa couplings arise in the same way, by introducing $N_d$ vectorlike down quarks $D_i$ and replacing the second and third line of Eq.~(\ref{eq:model}).  In the lepton sector, we introduce $N_\ell$ vectorlike doublets $L_i$, $N_e$ vectorlike charged singlets $E_i$, and $N_\nu$ vectorlike neutral singlets $\mathcal N_i$. 
Since our mechanism implies that the largest Yukawa coupling is of $\mathcal O(1)$, we introduce heavy Majorana masses for the fields $\mathcal N_{N_\nu+1}\equiv\nu_R$, implementing the sea-saw mechanism \cite{GellMann:1980vs,Yanagida:1979as,Mohapatra:1979ia}. 
The six integer numbers $N_q,\ N_u,\ N_d,\ N_e$ and $N_\nu$ are the only parameters that we treat non-stochastically, They can be viewed as the analogue of the FN charges in our model.

We have simulated this model, using flat priors with
\be
-m<M^i_{ab}<m\,,\qquad -m<K^i_{ab}<m\,,\qquad -1<Y^0_{ij}<1\,,
\label{eq:priors}
\ee
where $m$ is a heavy mass scale. It is clear that the physical Yukawa couplings (being dimensionless parameters) cannot depend on the mass $m$ and hence we will work in units of $m=1$, and analogous expressions hold for the down-quark and charged lepton sectors.
Note that the dependence on the variance of the prior then only enters in the quantity $Y^0$, and hence does not get magnified with powers of $N$ as would be the case in the simple toy model of the previous section. 

\begin{figure}
\centering
\includegraphics[width=8cm]{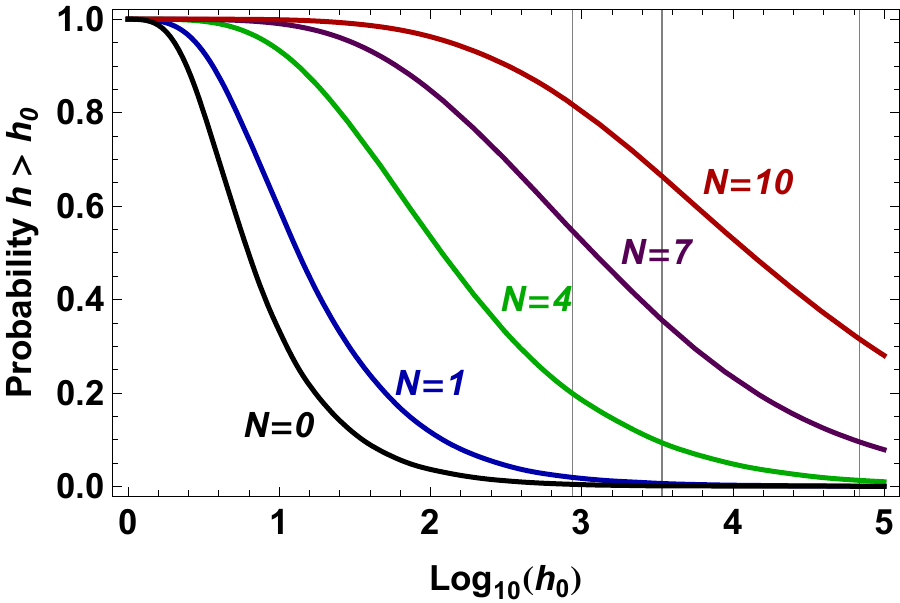}
\caption{Probabilities to find hierarchies greater than $h_0$ for various $N$. The vertical lines mark the SM values $h_d$, $h_e$, and $h_u$ respectively. }
\label{fig:cdf}
\end{figure}

\begin{figure}
\centering
\includegraphics[width=8cm]{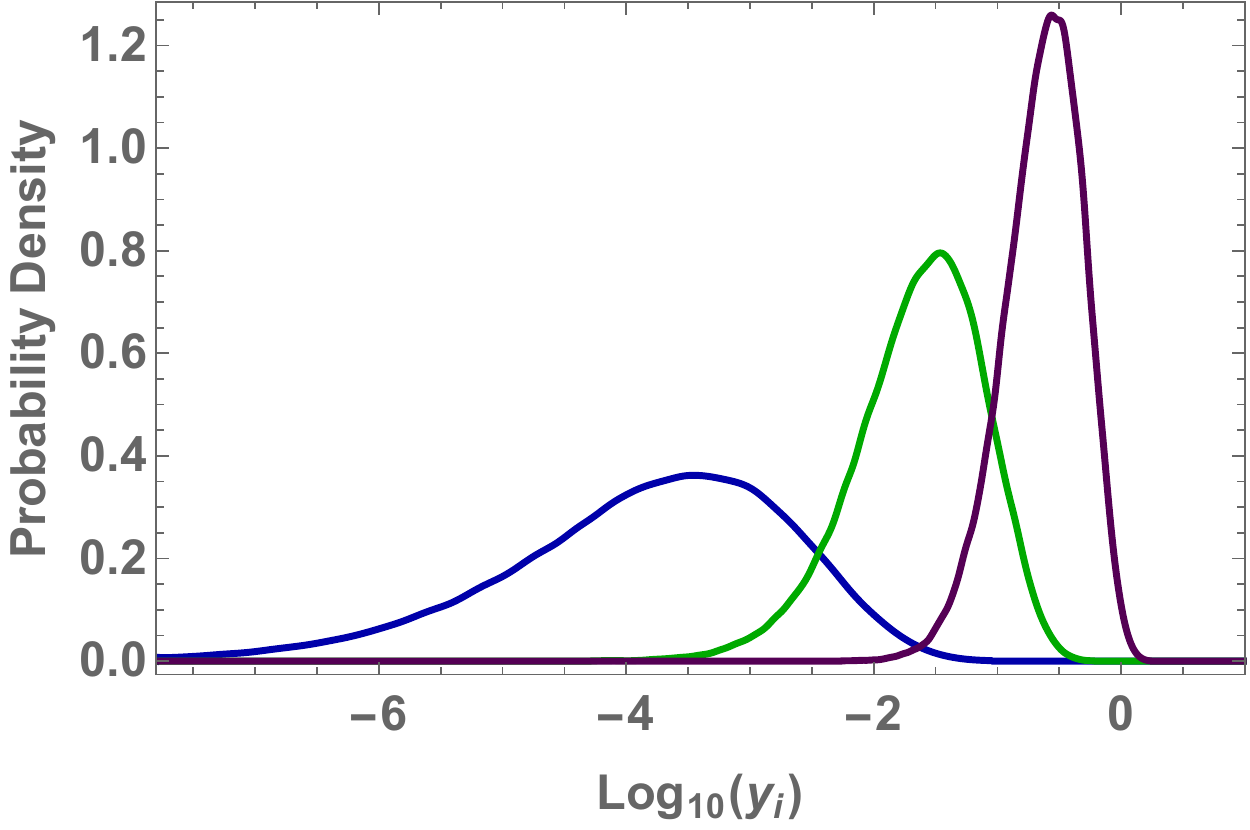}
\caption{Distribution of the small, medium, and large eigenvalues for $N=7$, for the full model defined in Eq.~(\ref{eq:model}).}
\label{fig:fullmodel}
\end{figure}

We will first consider the hierarchies $h$ defined in Eq.~(\ref{eq:hdef}), 
where the $y_i$ are now the physical eigenvalues, determined from 
\be
\det \left[\tilde Y^{u\,\dagger} (Z^q)^{-1}\tilde Y^u-y_i^2 Z^{u}\right]=0\,,
\ee
where $Z^{q,u}=1+\delta Z^{q.u}_N$ and $\tilde Y^u=(\tilde Y^q_{1}\cdots \tilde Y^q_{N_q})^\dagger Y^u_0\tilde Y^u_1\cdots\tilde Y^u_{N_u}$.
We will use as benchmarks  the SM hierarchies
\be
h_d\equiv 8.7 \times 10^2\,,\qquad h_e\equiv 3.4\times 10^3\,,\qquad h_u\equiv 6.9\times 10^4\,.
\label{eq:smh}
\ee
In Fig.~\ref{fig:cdf} we plot the probabilities for the occurence of hierarchies greater than $h_0$  as a function of $h_0$. 
To a good approximation, the eigenvalue distributions only depend on the sums ($N_q+N_u$, $N_\ell+N_e$ etc.), so for simplicity we report only the results for these sums (called $N$ in the following).
From the curve $N=0$, corresponding to the SM, one can see that even though $h>10$ occurs with probability of roughly $1/3$,  larger hierarchies are very unlikely. This however changes drastically when a few vectorlike fermions are introduced. 
As is evident from the curves, rather large  hierachies quickly become the rule rather than the exception.
In Tab.~\ref{tab:hier1} we quote explicitely the probabilities for the SM model benchmarks.
We also plot in Fig.~\ref{fig:fullmodel} the distributions for the actual eigenvalues in the case $N=7$. They do not look very different from the distributions of the simple product struture obtained in the previous section (see Fig.~\ref{fig:toymodel}). 
The independence of the exact form of the matrix product shows the robustness of our mechanism.

\begin{table}
\begin{center}
\begin{tabular}{cccc}
\toprule
$N$ & $p(h>h_d)$ & $p(h>h_e)$ & $p(h>h_u)$\\
\midrule
0 & $3.9\times 10^{-3}$ & $1.0\times 10^{-3}$ & $5.4\times 10^{-5}$ \\
1 & $2.0\times 10^{-2}$ & $5.9\times 10^{-3}$ & $3.6\times 10^{-4}$ \\
4 & 0.20 & 0.10 & $1.3 \times 10^{-2}$\\
7 &  0.55  &  0.36     & 0.097
\\
\midrule
1 & $9.1\times 10^{-2}$ & $3.4\times 10^{-2}$ & $3.2\times 10^{-3}$ \\
2 & 0.33 & 0.17 & $3.0 \times 10^{-2}$\\
4 &  0.80  &  0.63     & 0.28\\
\bottomrule
\end{tabular}
\end{center}
\caption{Probabilities for the hierarchies defined in Eq.~(\ref{eq:smh}) to occur by pure chance for various $N$. Upper block: priors chosen according to Eq.~(\ref{eq:priors}). Lower block: Modified prior Eq.~(\ref{eq:priorB}) with $q\ll1$.}
\label{tab:hier1}
\end{table}

\begin{figure}
\centering
\includegraphics[width=8cm]{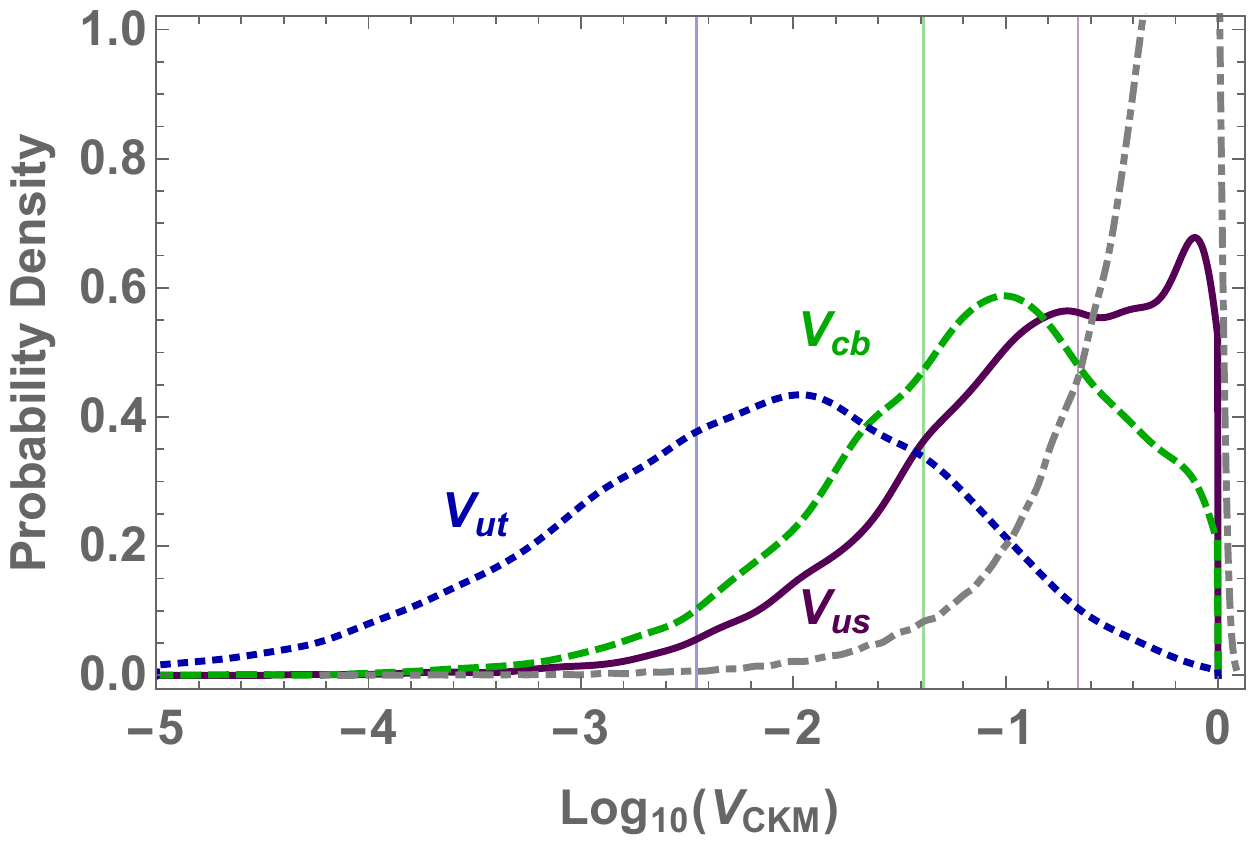}
\caption{Distribution of CKM matrix elements for $N_q=7$, $N_d=1$, $N_u=4$. The experimental values are indicated as vertical lines. For comparison, we show the case $N_q=0$, $N_d=5$, $N_u=11$, in which all angles follow the same distribution (dashed-dotted line).}
\label{fig:CKM}
\end{figure}

The CKM mixing angles are well reproduced from $N_q\approx 6-8$ (see Fig.~\ref{fig:CKM}) while 
absence of alignment between the charged leptons and neutrinos require $N_\ell\approx 0-1$. 
One can also ask the question whether one can find a choice of parameters that is compatible with $SU(5)$ quantum numbers of a potential grand unified theory.
Taking into account both the masses and mixings, we find that 
\be
N_q=N_u=N_e=6\,,\qquad N_\ell=N_d=0
\ee
works rather well.
Furthermore, in order to avoid a too-large hierarchy between the two heaviest neutrinos, we choose $N_\nu\lesssim 2$. 
The case $N_\ell=N_\nu=0$ then corresponds to standard anarchic see-saw neutrinos. Taking $N_\nu\neq 0$ in addition creates a small hierarchy in the Neutrino Yukawa couplings, but as long as $N_\ell=0$ we do not generate any alignment with the charged lepton sector,  and mixing angles will stay large.

One could be worried that a large number of new fermions  will give rise to Landau poles for the gauge couplings in the UV. 
Notice however that up to now nothing has been said about the mass scale for the new fermions.
It is fully consistent (and in fact natural) that these masses are at or just below the Planck scale. In this case the model is consistent with the absence of any additional New Physics below the Planck scale.  
Notice that such a high mass scale automatially suppresses dangerous FCNCs which would otherwise be an issue due to the presence of the vector like quarks and leptons. 
An alternative scenario would be an additional UV completion of the model at a lower scale, in which case a detailed assessment of  flavor violating effects would be necessary.

Finally, we would like to point out two more  ways to even further improve the performance of the mechanism. Firstly, assume that there is a reason for the mass mixings $K_i$ to be systematically suppressed with respect to the masses $M_i$. 
We can incorporate this assumption easily by modifying the priors for $K_i$ as 
\be
-q\, m<K^i_{ab}<q\,m\,,
\label{eq:priorB}
\ee
with $q$ a dimensionless number $q\ll 1$. In this case, one has approximately $\delta Z_k\approx 0$, and
\be
\tilde Y^\ms_k\approx M_k^{-1}K^\ms_k\,.
\ee
Thus in this approximation of small $q$ the physical Yukawa couplings simply scale as $q^N$ and the hierarchies, being ratios of eigenvalues, become independent of $q$. The probabilities from these distributions are  also given in Tab.~\ref{tab:hier1}.
We find that hierarchies are generated even more efficiently, presumably because accidentally small vectorlike masses have a larger effect or occur more common than in the case $q=1$ considered above.  

A second possible modification is the assumption of some kind of minimal flavor violation meachanism, that fixes all mixings to be proportional to, say, $Y_0$,
\be
K_i\propto Y_0\,.
\ee
For simplicity we may assume that the masses $M_i$ are proportional to the identity. There are always mild hierarchies $h_0$ in the random eigenvalues of $Y_0$ (confer the $N=0$ curve in Fig.~\ref{fig:cdf}), and since there is a basis in which everything is diagonal simultaneously, these hierarchies are coherently amplified $h\sim h_0^N$. We leave the construction of an explicit mechanism to future work.

\section{Conclusions}

In this note we have presented a new explanation for the large hierarchy of fermion masses present in the SM. It is based on the observation that products of a few random matrices typically feature strong hierarchies in their eigenvalue spectrum, even though the individual entries are of order unity. 
Moreover, we have shown that the peculiar form of the CKM matrix  can easily be recovered within this paradigm, as up and down Yukawa couplings can become naturally aligned if they include common factors as in Eq.~(\ref{eq:CKMmodel}). 
We have presented a model that generates an effective Yukawa coupling as a product of several matrices. This model is by no means unique, and we expect that any model with such a product structure has similar eigenvalue distributions. 
Finally, even though we have focused on the case of real random matrices for simplicity, CP violation can easily be accommodated by 
 including random complex phases.

\section*{Acknowledgements}
I would like to thank Arman Esmaili for useful discussions and comments on the manuscript, and the  Conselho Nacional de Desenvolvimento Cient\'ifico e Tecnol\'ogico (CNPq) for support under fellowship number 307536/2016-5. 

\bibliography{paper}
\bibliographystyle{hieeetr}

 
\end{document}